\documentclass[fleqn,usenatbib]{mnras}

\usepackage{newtxtext,newtxmath}
\usepackage[T1]{fontenc}

\DeclareRobustCommand{\VAN}[3]{#2}
\let\VANthebibliography\thebibliography
\def\thebibliography{\DeclareRobustCommand{\VAN}[3]{##3}\VANthebibliography}

\usepackage{graphicx}	% Including figure files
\usepackage{amsmath}	% Advanced maths commands
\usepackage{ulem}
\title[Subaru constraints on unbound objects]{Constraints on sub-terrestrial free-floating planets from Subaru microlensing observations}

\author[W. DeRocco et al.]{
William DeRocco,$^{1,2}$\thanks{E-mail: wderocco@ucsc.edu}
Nolan Smyth,$^{1,2}$
and Stefano Profumo$^{1,2}$
\\
$^{1}$Department of Physics, University of California, Santa Cruz (UCSC),
Santa Cruz, CA 95064, USA\\
$^{2}$Santa Cruz Institute for Particle Physics (SCIPP),
Santa Cruz, CA 95064, USA}

\date{Accepted XXX. Received YYY; in original form ZZZ}

\pubyear{2023}

\begin{document}
\label{firstpage}
\pagerange{\pageref{firstpage}--\pageref{lastpage}}
\maketitle

\begin{abstract}
The abundance of protoplanetary bodies ejected from their parent star system is presently poorly-constrained. With only two existing optical observations of interstellar objects in the $10^{8} - 10^{10}$ kg mass range and a small number of robust microlensing observations of free-floating planets (FFPs) in the $10^{24} - 10^{25}$ kg mass range, there is a large range of masses for which there are no existing measurements of the unbound population. The three primary microlensing surveys currently searching for FFPs operate at a cadence greater than 15 minutes, which limits their ability to observe events associated with bodies with a mass much below an Earth mass. We demonstrate that existing high-cadence observations of M31 with the Subaru Hyper Suprime-Cam place constraints on the abundance of unbound objects at sub-terrestrial masses, with peak sensitivity at $10^{-4}~M_\oplus$ for Milky Way lenses and $10^{-1}~M_\oplus$ for lenses in M31. For a fiducial $\frac{dn}{dM}\propto M^{-2}$ mass distribution, we find that the abundance of unbound objects is constrained to $n_\text{unbound} < 1.4 \times 10^{7} ~\rm{pc}^{-3}$ for masses within 1 dex of $10^{-4}~M_\oplus$. Additionally, we compute limits on an artificial ``monochromatic'' distribution of unbound objects and compare to existing literature, demonstrating that the assumed spatial distribution of lenses has very significant consequences for the sensitivity of microlensing surveys. While the observations ultimately do not probe abundances suggested by current models of planetary formation, our limits place direct observational constraints on the unbound population in the sub-terrestrial mass range and motivate new observational strategies for microlensing surveys.
\end{abstract}


%%%%%%%%%%%%%%%%%%%%%%%%%%%%%%%%%%%%%%%%%%%%%%%%%%

%%%%%%%%%%%%%%%%% BODY OF PAPER %%%%%%%%%%%%%%%%%%

\section{Introduction}
\label{sec:introduction}

During the early stages of planet formation, chaotic dynamical interactions between planetesimals and their environment are expected to eject the majority of the mass contained in the bound planetesimal reservoir, producing a large population of macroscopic unbound objects (UBOs), ranging in mass from less than a kilogram to well above the mass of Earth. Despite strong theoretical motivation for the existence of this population, only in recent years have observations begun to actively explore it\footnote{Note that in some sense, \textit{microscopic} UBOs were observed earlier via {\it in situ} measurements of interstellar dust in the Solar System \citep{2019SSRv..215...43S}.}. Observations have occurred primarily in two mass ranges. At masses between $\approx 10^{8} - 10^{10}$ kg \citep{Seligman_2020,Hui_2020}, optical surveys have observed two interstellar objects (ISOs), roughly comet-mass bodies that have been ejected from their host star and subsequently transited the Solar System. Anomalous features of the two observed ISOs, 1I/'Oumuamua \citep{2017Natur.552..378M} and 2I/Borisov \citep{2021SoSyR..55..124B}, have stimulated vigorous debate, and motivate better characterizing the UBO population\footnote{At even lower masses, there also exist putative detections of interstellar meteors \citep{siraj20222019}, however these are subject to debate \citep{2022JIMO...50..140V,brown2023proposed}.}. These efforts are complemented at much higher masses ($\gtrsim 10^{24}$ kg) by multi-year high-cadence microlensing surveys that search for the magnification of background stars due to gravitational lensing by non-luminous objects. Since the lensing effect is purely gravitational, this technique is most sensitive to high-mass lenses. Only recently have technological advancements in high-cadence imaging allowed these surveys to probe the terrestrial-mass range, with multiple collaborations reporting the first observations of terrestrial-mass ``free-floating planets'' (FFPs\footnote{This term is often also taken to encompass Jupiter-mass failed brown dwarfs that form {\it in situ}. In this paper, we focus only on lower-mass FFPs as these are expected to have been ejected from star systems, hence represent the highest-mass constituents of the UBO population.}) \citep{2019A&A...622A.201M,Mroz_2020,koshimoto2023terrestrial,sumi2023freefloating}. At present, there are only $\approx 3$ events for which the light-curve enables a mass estimate placing it in the terrestrial range. 

Between the ISO and FFP observations, there are over ten orders of magnitude in mass where the unbound population is poorly constrained. This lack of data on UBOs strongly motivates exploring all possible avenues to improve our understanding both within the ISO and FFP mass ranges as well as the unexplored gulf between them. To this end, here we carefully re-purpose existing observations by the Subaru telescope \citep{niikura_microlensing_2019} to place constraints on the abundance of UBOs in the mass range $\approx 10^{-5}~M_\oplus-1~M_\oplus$.

The remainder of this paper is organized as follows:  In section \ref{sec:microlensing}, we outline the formalism of microlensing and discuss existing microlensing surveys. In section \ref{sec:ISOs}, we discuss the theoretical distribution of UBOs and compare to existing data. We present a brief summary of the observations performed by the Subaru Hyper Suprime-Cam (HSC) in section \ref{sec:subaru}. In section \ref{sec:results}, we derive limits from the HSC microlensing survey, and provide constraints on the UBO population. We then connect this finding to estimates from existing ISO and FFP data and conclude with a discussion of the implications of our results in
section \ref{sec:discussion}. Finally, appendix \ref{app:scalings} discusses the scaling behavior of the sensitivity curve at high and low lens masses.

\section{Microlensing}
\label{sec:microlensing}

Gravitational lensing of astrophysical objects is a powerful technique for observing massive, non-luminous objects \citep{paczynski_gravitational_1986}. In this section, we introduce the formalism associated with this technique and discuss existing surveys with sensitivity to the Galactic unbound population. 

\subsection{Formalism}
\label{sec:formalism}

High-mass lenses produce multiple images of a background source star that can be used to characterize the mass and morphology of the lens. At typical planetary masses, however, the multiple images formed through gravitational lensing are not resolved; rather, the effect manifests as a magnification of the background star (``source'') by a factor of $A = \frac{\phi}{\phi_0}$
where $\phi_0$ is the flux in the absence of lensing. This effect is known as \textit{microlensing}. The typical distance scale  associated to the lensing produced by an object of mass $M$ is the Einstein radius, $R_E$, defined as 
\begin{equation}
\label{eq:RE}
    R_E = \sqrt{\frac{4GM d_L(1- d_L/d_S)}{c^2}},
\end{equation}
where $d_S$ and $d_L$ represent the distances between the observer and the source, and the observer and the lens, respectively. The angular size of the Einstein radius is therefore $\theta_E = \frac{R_E}{d_L}$. The typical timescale for a microlensing event is given by the time for the source to cross the Einstein ring in the plane of the lens. This ``Einstein crossing time'' is defined as 
\begin{equation}
\label{eq:tE}
    t_E = \frac{\theta_E}{\mu_\text{rel}} \approx 4.1~\text{hr} \left(\frac{M}{M_\oplus}\right)^{1/2}\left(\frac{d_L}{\text{kpc}}\right)^{1/2}\left(\frac{v_T}{50~\text{km/s}}\right)^{-1},
\end{equation}
where $v_T$ is the transverse velocity of the lens and we have assumed $d_L \ll d_S$ in the final expression.

In the geometric optics approximation, the magnification for a point source is \citep{nakamura_wave_1999}

\begin{equation}
    A_{\rm geo} = \frac{u^2 + 2}{u\sqrt{u^2 + 4}},
    \label{eq:Ageo}
\end{equation}
where $ u $ is the impact parameter in the lens plane in units of the Einstein radius. Note that setting $u = 1$ corresponds to $A_{\rm geo} = 1.34$, the conventional threshold for a detectable event.
This expression breaks down when the angular size of the source becomes comparable to that of the lens. In this regime, finite-source effects reduce the peak magnification since the lens magnifies only part of the source at any time. While the peak magnification of a source is technically infinite in the point-source regime, it remains finite in the finite-source regime.

The parameter 
\begin{equation}
    \rho \equiv \frac{\theta_S}{\theta_E} = \frac{R_S/d_S}{R_E/d_L},
\end{equation}
where $\theta_S$ is the angular size of the source, quantifies the extent to which finite-source effects are important. For $\rho \ll 1$, the point-source limit holds to a good approximation. When $\rho \gtrsim 1$, this approximation breaks down and the net magnification is instead calculated using the average magnification over the extent of the star. 

The average magnification is calculated by performing an integral in circular coordinates $(r,\theta)$ over the source disk, the center of which is defined as the origin. Due to the symmetry of the system, we take the center of the lens to be located a distance $u$ away from the center of the source along $\theta = 0$. This yields \citep{matsunaga_finite_2006, witt_can_1994, sugiyama_revisiting_2019}
\begin{multline}
    A_{\mathrm{finite}}(u,\rho) \equiv \\
    \frac{1}{\pi \rho^2} \int_0^{\rho} dr \int_0^{2\pi} d\theta ~r ~A_{\rm{geo}}\Big( \sqrt{r^2 + u^2 - 2ur \cos(\theta)} \Big)
    \label{eq:Afinite}
\end{multline}

As a result, the threshold impact parameter corresponding to a detection deviates from the geometric optics case. We adopt the methodology proposed in \citet{sugiyama_revisiting_2019, smyth_updated_2020} and solve for $u_{\rm{T}}$, the maximum impact parameter that results in a detectable event ($A_{\mathrm{finite}}(u_{\rm{T}},\rho) = 1.34$). This defines the phase space for the calculation of the expected event rate defined in equation~\ref{eq:difRate}.

While finite-source effects serve to diminish the magnification of the source, making such events more difficult to detect, they also introduce features to the light-curve that can be used to measure $\theta_E$. For a typical microlensing event, the only observable quantity is $t_E$. Unfortunately, due to the degeneracy of lens mass, distance, and transverse velocity in the expression for $t_E$ (equation \ref{eq:tE}), a measurement of $t_E$ is often insufficient to characterize the nature of the lens. A measurement of $\theta_E$ from finite-source effects partially breaks this degeneracy, hence such ``finite-source point-lens'' (FSPL) events allow for a mass estimate.\footnote{Estimating a mass requires an assumption about $d_L$, hence existing FSPL observations of putative terrestrial-mass FFPs for Bulge-oriented surveys often quote two values for the mass, one for an FFP in the Galactic bulge and one for an FFP in the Galactic disk.}

\subsection{High-cadence surveys}

There are three primary collaborations performing dedicated multi-year high-cadence microlensing surveys. These are the Optical Gravitational Lensing Experiment (OGLE-IV)~\citep{udalski2015ogleiv}, the Korea Microlensing Telescope Network (KMTNet)~\citep{2016JKAS...49...37K}, and Microlensing Observations in Astrophysics (MOA-II)~\citep{Abe:2008rpq}. All three of these surveys are optimized for the detection of microlensing events; they have wide fields of view focused on areas with high stellar density, like the Galactic bulge, and observe at a high cadence for durations spanning years. Here, ``high cadence'' means observations taken roughly more than once per night. Yet, recent advancements have enabled these collaborations to achieve cadences as rapid as 15 minutes. Considering the $\mathcal{O}(\text{hour})$-long timescales of microlensing events for terrestrial masses (see equation \ref{eq:tE}), it is only recently that these collaborations have begun to observe microlensing events consistent with a population of low-mass free-floating planets. As discussed in section~\ref{sec:formalism}, finite-source point-lens (FSPL) events allow mass estimates of putative FFP lenses. At present there are only three FSPL events consistent with terrestrial-mass FFPs.\footnote{These are OGLE-2012-BLG-1323 \citep{2019A&A...622A.201M}, OGLE-2016-BLG-1928 \citep{Mroz_2020}, and MOA-9y-5919 \citep{koshimoto2023terrestrial}.}

With typical cadences of roughly an hour for most fields, these collaborations are only marginally sensitive to lenses in the terrestrial mass range. Due to the $t_E \propto M^{1/2}$ dependence of the crossing time on lens mass, faster cadences afford the opportunity to probe lower masses. Here, we use observations performed by the Subaru Hyper Suprime-Cam (HSC) with a cadence of 2 minutes to exploit this fact. This is roughly one order of magnitude faster than the fastest operational mode for the microlensing surveys listed above; due both to this faster cadence and the lower velocity dispersion of disk lenses (see app. \ref{app:scalings}), these observations are sensitive to lens masses roughly four orders of magnitude below those observable by the existing microlensing surveys, providing strong motivation to use the HSC observations as a probe of sub-terrestrial UBOs.

\section{The unbound population}
\label{sec:ISOs}

In this section, we review what is currently known about the abundance of unbound objects and state the assumptions on the UBO mass distribution that we employ in our analysis.

\subsection{Mass distribution of unbound objects}
\label{sec:massdist}

The UBO population is thought to be predominantly composed of planetesimals of varying mass that were ejected from their birth system during the chaotic early phases of system formation. There are various processes that can lead to the release of a planetesimal, including release during disk dispersal, removal by nearby stars in the birth cluster, gravitational scattering by planets, slow drift from exo-Oort clouds, ejection during post-red giant mass loss, and ejection by an inner binary star system \citep{fitzsimmons2023interstellar}. Since all these processes draw from a star's existing planetesimal reservoir, the abundance of UBOs depends strongly on this reservoir's properties.

Characterizing these reservoirs is a major research focus in the exoplanet community as it bears heavily on the ultimate distribution of planets formed within a system.  At present, the distribution of planetesimals is still largely unknown. It is often modeled as a power law, either in radius or mass. Here we adopt the form 
\begin{equation}
\label{eq:mass_func}
\frac{dN}{d \log_{10} (M)} = \mathcal{N} \Big( \frac{M}{M_{\rm{norm}}}\Big)^{-p}
\end{equation}
where $\mathcal{N}$ is the total number of UBOs per star at mass $M$ scaled by a  normalization mass $M_{\text{norm}}$. Throughout the rest of the paper, we take $M_{\rm{norm}} = M_{\oplus}$ and all logarithms to be base 10. For ISOs, the distribution is often defined in terms of the radius $r$ of the ISO as
\begin{equation}
    \frac{dN}{dr} \propto r^{-(\alpha + 1)},
\end{equation}
where $\alpha = 3p$.\footnote{Note that this is rewritten as $N(r) \propto r^{-\alpha}$ in the literature, where $N(r)$ is taken to be the number above a radius $r$, see e.g. \citet{Engelhardt_2017}.} Many estimates have been made on the normalization and power-law index of this distribution, both from results of simulations \citep{1968IAUS...33..486D,2009ApJ...704..733M,G_sp_r_2012} and from observational data \citep{Strigari_2012,sumi2023freefloating,10.5303/JKAS.2022.55.5.173,2019arXiv190603270S,Landgraf_2000} that span a range $\approx 0.66 - 1.33$, with $p \approx 1.0$ often assumed as the fiducial value.\footnote{In the context of ISOs, the index $p = 0.83$ often also appears as a fiducial value, as it corresponds with the theoretical value given by a self-similar collisional cascade \citep{1968IAUS...33..486D}, however more sophisticated models have yielded higher estimates of $p$ \citep{G_sp_r_2012}.} 

How well a power-law describes the full UBO population over both ISO through FFP mass ranges is an open question. Physical processes occurring at different scales may induce features in the distribution. For example, efficient gravitational reaccumulation of small fragments onto larger planetesimals would result in an overabundance of high mass planetesimals ($\gtrsim 10^{18}$ kg) at the expense of a relative underabundance at low masses ($\lesssim 10^{12}$ kg) \citep{Lohne_2008}. Despite this, numerical studies have found that the mass distribution in debris disks is well-approximated by a power law over many orders of magnitude \citep{G_sp_r_2012}. Hence, to compare with existing literature, we adopt a power-law form for the mass distribution and allow $p$ to vary between 0.66 and 1.33.

\subsection{Abundance}
\label{sec:abundance}

Prior to the recent ISO and FFP observations, estimates of the abundance of UBOs could only be derived from theory. With observations now made in both mass ranges, data-driven estimates of the abundance of UBOs can be made.

In the ISO mass range, the observations of 1I/`Oumuamua and 2I/Borisov have produced estimates of the density of UBOs of size $> 100$ meters of $\approx 0.1 - 0.2~\text{au}^{-3}$ ($8.8\times 10^{14}-1.8\times10^{15}~\text{pc}^{-3}$) \citep{Do_2018,Jewitt_2017,Trilling_2017}. In terms of mass, this corresponds to a limit on UBOs with masses $\gtrsim 2\times10^{9}$ kg ($r \gtrsim 100$ m for typical cometary densities of $0.5~\mathrm{g/cm^3}$).\footnote{Note that this estimate is an extrapolation for masses well-above $2\times 10^{9}$ kg, however due to the steep mass-dependence of the UBO distribution, the dominant contribution to the abundance arises from the lowest masses. It is therefore approximately correct to quote the limit as applying to all masses $>2\times 10^{9}$ kg as \citet{Do_2018} do.}
Taking the ISO population to be well-approximated by a collisional cascade, as was done in \citet{Jennings_2020}, yields a rough estimate for the UBO mass distribution of 
\begin{equation}
\label{eq:dohnanyi}
    \left(\frac{dN}{d\log M}\right)_\text{ISO+cascade} \approx \left(\frac{1.5\times10^{16}}{\text{star}}\right) \left(\frac{M}{2\times 10^9\,\text{kg}}\right)^{-0.83},
\end{equation}
where we have taken the local stellar density to be $\approx0.1\, \text{pc}^{-3}~$ \citep{Golovin_2023}.\footnote{Note that extrapolating this distribution to planetary masses yields estimates of thousands of Earth-mass FFPs per star, which is not well-supported by data. This indicates that a simple collisional cascade model is no longer realistic for high-mass UBOs.}

As discussed in section~\ref{sec:microlensing}, for the FFP mass range relevant to the sub-terrestrial masses discussed in this paper, recent observations by existing microlensing surveys of the first terrestrial-mass free-floating planets \citep{2019A&A...622A.201M,Mroz_2020,koshimoto2023terrestrial} have allowed estimates of the abundance of high-mass UBOs.\footnote{Quasar microlensing studies \citep{Dai_2018,Bhatiani_2019} also suggest a possible FFP abundance of $\sim 100$ Moon-mass to Jupiter-mass objects per star for an assumed $p=1$ power law, which is broadly consistent with our fiducial curve.}
The MOA collaboration estimates a UBO distribution given by \citep{sumi2023freefloating}
\begin{equation}
\label{eq:moa}
    \left(\frac{dN}{d\log M}\right)_\text{MOA} = \left(\frac{2.18^{+0.52}_{-1.40}}{\text{star}}\right) \left(\frac{M}{8\,M_\oplus}\right)^{-(0.96^{+0.47}_{-0.27})},
\end{equation}
based upon FFP observations alone while KMTNet chooses to include ISO measurements  to estimate \citep{10.5303/JKAS.2022.55.5.173}
\begin{equation}
\label{eq:kmt}
    \left(\frac{dN}{d\log M}\right)_\text{KMT+ISO} = \left(\frac{0.4^{+0.2}_{-0.2}}{\text{star}}\right) \left(\frac{M}{38\,M_\oplus}\right)^{-(0.92^{+0.06}_{-0.06})}.
\end{equation}
The inclusion of ISO data to constrain $p$ implicitly assumes that a single power-law is a good approximation of the UBO distribution over the $\approx 17$ orders of magnitude in mass separating ISOs from FFPs; a more conservative estimate of $p$ using purely the KMTNet FFP observations yields $0.9 \lesssim p \lesssim 1.2$ \citep{10.5303/JKAS.2022.55.5.173}.
Both collaborations have set their reference masses $M_\text{norm}$ to be the approximate mass to which their analysis is most sensitive. As a result, their estimates are only supported by data for masses near that value. 
The sensitivity to lower masses is cut off by the cadence of observation, as (in the limit of no finite-source effects) the Einstein crossing time scales with $M_\text{lens}^{1/2}$. Note that due to the approximate $dN/dM \propto M^{-2}$ dependence of the UBO mass function, reducing the cadence by a single order of magnitude allows detection of masses two orders of magnitude lower; these are expected to be roughly four orders of magnitude more abundant. (See appendix \ref{app:scalings} for a discussion of the cadence scaling of a survey's peak mass sensitivity.) This is why exploring this mass window does not require a multi-year observational campaign such as those conducted by OGLE, KMTNet, and MOA. 

Synthesizing all existing data and simulation results (see section~\ref{sec:massdist}), one can construct a reasonable approximate fiducial mass function as
\begin{equation}
\label{eq:fid}
    \left(\frac{dN}{d\log M}\right)_\text{fid} = \left(\frac{10}{\text{star}}\right) \left(\frac{M}{M_\oplus}\right)^{-1}.
\end{equation}
It is worth noting that given the current data, \textit{unbound} terrestrial-mass planets are expected to outnumber bound terrestrial planets, providing further motivation to explore this population as a probe of planetary dynamics.

\section{Subaru/HSC observations}
\label{sec:subaru}

On the night of November 23, 2014, the Subaru Hyper Suprime-Cam  performed an observation of M31 at a cadence of 2 minutes to search for short-duration microlensing events \citep{niikura_microlensing_2019}. Within the 7-hour observation, only one candidate event was detected. The results of this observation were used to place significant constraints on the abundance of primordial black holes (PBHs) in the mass range $10^{19} - 10^{23}$ kg \citep{niikura_microlensing_2019, smyth_updated_2020, sugiyama_revisiting_2019, croon_subaru_2020}.

\begin{table}
\centering
\begin{tabular}{|l|c|}
\hline
Instrument                  & Subaru Hyper-Suprime-Cam                             \\ \hline
Survey Date                 & November 23, 2014                                 \\ \hline
Target                      & M31                                   \\ \hline
Galactic Longitude \( l \) (degrees)           & 121.2                                 \\ \hline
Galactic Latitude \( b \) (degrees)           & $-21.6$                           \\ \hline
Field of View Diameter (degrees)   & 1.5                                 \\ \hline
Source Distance (kpc)              & 770                                   \\ \hline
Cadence (minutes)           & 2                                    \\ \hline
Observation Time (hours)    & 7                                     \\ \hline
\( N_{\text{sources}}\) (millions)    & \( 87 \)                  \\ \hline
\end{tabular}
\caption{Summary of parameters for the Subaru HSC M31 survey.}
\label{tab:parameters}
\end{table}

The resulting limits apply not only to PBHs, but to UBOs as well.\footnote{This point was made prior to our work by \citet{Jennings_2020}, however a rigorous analysis to compute the limit was not performed.} As is clear from the discussion in section~\ref{sec:microlensing}, the 2-minute cadence allows these observations to constrain the abundance of UBOs even at sub-terrestrial masses. There are, however,  a number of key differences between a PBH population and UBO population of lenses that must be taken into account when attempting to constrain the UBO population. 

\begin{figure}
	\centering
	\includegraphics[width=\columnwidth]
 {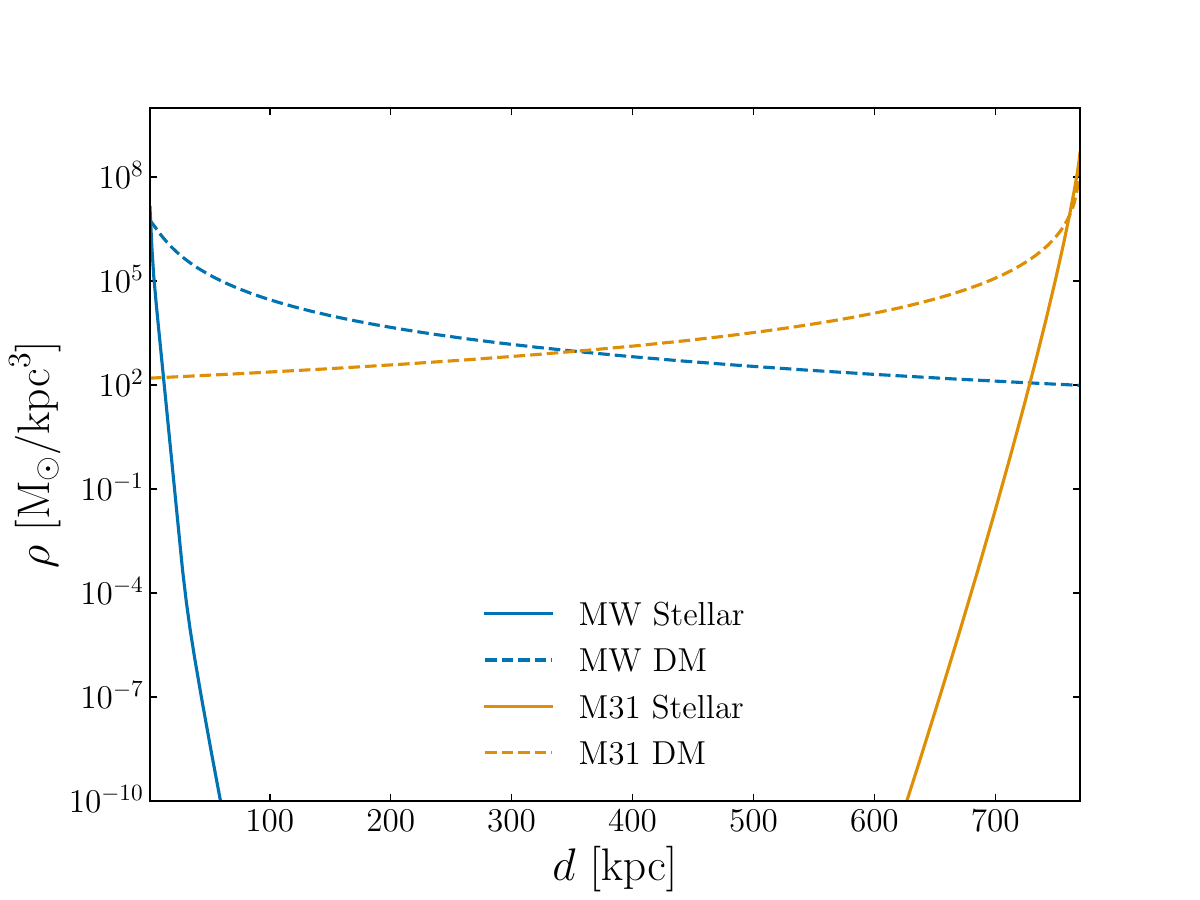}
	\caption{The dark matter (DM) and stellar density from the Milky Way (MW) and M31 along the line-of-sight toward M31. Note that the UBO density is expected to be proportional to the stellar density. The contribution from DM is significant over all distances due to the large extent of both halos. This makes M31 a prime candidate for dark matter lensing surveys. For UBOs, the contribution from MW lenses is minimal since the stellar density decreases beyond the scale height of the Milky Way thin disk ($\approx 0.3~$kpc) and the contribution from M31 lenses is largest close to the sources, where finite source effects are most significant.}
	\label{fig:line_of_sight}
\end{figure}

Firstly, UBOs do not follow the same spatial distribution as PBHs: PBHs are assumed to form spherical halos surrounding the Milky Way and M31; In contrast, the UBO population is expected to trace the stellar density within each galaxy. This is because UBOs are generally ejected with low velocities and therefore do not travel a significant distance from their parent star system. It is for this reason that existing microlensing surveys tend to observe the Galactic bulge, as this line-of-sight has a large density of UBO lenses.
The 2014 Subaru observation chose M31 as an ideal target specifically due to the large integrated line-of-sight density of \textit{PBH lenses} in the Milky Way's dark matter halo. Since the Earth-M31 line-of-sight is oriented out of the plane of the Milky Way, the relative integrated density of \textit{UBO lenses} is much smaller, as the stellar density decreases rapidly beyond the scale height of the Milky Way thin disk ($\approx 0.3~$kpc~\citep{doi:10.1146/annurev-astro-081915-023441}). This can be seen in Fig.~\ref{fig:line_of_sight}, which shows both the stellar density (solid) and dark matter density (dashed) along the line of sight. It is perhaps unsurprising, given this figure, that the difference in spatial morphology significantly weakens the sensitivity of the Subaru observations to UBOs in comparison to PBHs.

A further difference in the UBO and PBH populations is that the dark matter halo of PBHs is virialized. As a result, PBH lenses typically move with velocities on the order of the Galactic virial velocity, $\approx 270$ km/s. UBOs are thought to be ejected with velocities $< 10$ km/s with respect to their parent star's frame of rest \citep{fitzsimmons2023interstellar}, hence tracking the stellar velocity dispersion, which is roughly an order of magnitude lower. For this reason, the timescale of UBO microlensing events is roughly an order of magnitude larger than that of PBHs for identical lens masses and distances, allowing greater sensitivity to lower lens masses for UBOs in than PBHs.

Finally, the analysis performed in \citet{niikura_microlensing_2019} assumed a monochromatic mass spectrum for PBHs since a narrow distribution of PBH masses is motivated from theoretical considerations \citep{Carr_2022}. UBOs, in contrast, are predicted to follow an approximate power-law distribution of masses, as described in section \ref{sec:massdist}. We therefore choose to set limits on the parameters of this power law rather than to adopt an unphysical monochromatic distribution of UBOs. We do, however, show a ``monochromatic'' limit in Fig.~\ref{fig:monochromatic_mass} to demonstrate the significant loss of sensitivity due to the low UBO density along the line-of-sight in comparison to the rescaling of the PBH bound that currently appears in the literature \citep{johnson_predictions_2020,Jennings_2020}. 

Note, finally, that the finite-size effects are particularly important for UBO lenses within M31, where the lens-source distance is small. This results in reduced sensitivity to masses below $\sim 10^{-2}~M_{\oplus}$. While the MW UBO population below this mass can still be probed, the line-of-sight morphology described above leads to a significant attenuation of the overall expected event rate for UBOs with masses lower than this, see Fig.~\ref{fig:diff_rate_M_M31} and the surrounding discussion.

\subsection{Expected Event Rate}
\label{sec:expected_event_rate}

Subaru is not able to resolve individual stars in M31, hence it operates in a ``pixel lensing'' regime \citep{Calchi_Novati_2009,niikura_microlensing_2019}. In this regime, microlensing events appear as increased flux in a given pixel. An individual pixel measures the combined flux of many stars. Operating in this regime immediately precludes the ability to extract the magnification $A$, as the baseline flux of the lensed star cannot be measured separately from the blended flux of all stars in the pixel. As a result, one cannot measure the physical Einstein crossing time, $t_E$. Instead, one considers the full-width-half-maximum (FWHM) event timescale\footnote{ In the absence of finite-source effects, $t_{\rm{FWHM}} \propto u_0 t_E$ when $u_0 \ll 1$ (see e.g. \citet{Calchi_Novati_2009}). In the finite-source regime, the event duration is determined primarily by the angular size of the source rather than by $\theta_E$. In pixel-lensing, $t_E$ cannot be measured directly in either case.}, which can be measured from the pixel flux light-curve directly and is defined via

\begin{equation}
    A\Big(\frac{t_{\rm{FWHM}}}{2}\Big) - 1 \equiv \frac{A(u_0) - 1}{2},
\end{equation}
where $u_0$ is the impact parameter at the point of closest approach between the lens and source. Without a measurement of the physical quantities $t_E$ and $\theta_E$, it is not possible to reconstruct the properties of the lens given the detection of a microlensing event. However, it is straightforward to calculate the expected rate of events as a function of the observed $t_{\rm{FWHM}}$.

The expected event rate calculation proceeds as follows: assuming an isotropic Maxwellian velocity distribution for the lenses, the expected differential rate of microlensing events for a single source per observation time is given by

\begin{multline}
    \label{eq:difRate}
     \frac{d\Gamma}{d t_{\rm{FWHM}}} = 2 \int_{\rm{M_{min}}}^{\rm{M_{max}}} dM \int_0^{d_s}dd_L \int_0^{u_T}du_0 \\
     \frac{1}{\sqrt{u_T^2 - u_0^2}} \frac{\rho_M}{M}\frac{v_T^4}{v_c^2} \exp \Big[ -\frac{v_T^2}{v_c^2}\Big] f(M) \varepsilon(t_{\rm{FWHM}}),
\end{multline}
where $f(M)$ is the probability distribution function of lens masses, $v_T = 2R_E \sqrt{u_{\rm{HM}}^2 - u_{\min}^2}/t_{\rm{FWHM}}$ is the transverse velocity of the lens, $\rho_M$ is the mass density of the lens population, $d_S$ is the distance to the source star, and $\varepsilon(t_{\rm{FWHM}})$ is the detection efficiency. $u_{\rm{HM}}$ is the impact parameter at which the half-maximum magnification is reached and can be solved for analytically in the point-source regime and numerically in the finite-source regime via Eqs. \ref{eq:Ageo} and \ref{eq:Afinite}, respectively.
There will be a contribution from populations in both the Milky Way and M31 to the total event rate, $\Gamma = \Gamma_{\rm{MW}} + \Gamma_{\rm{M31}}$. We must specify spatial and velocity distributions of UBOs in both the Milky Way and M31 to compute the event rate. In our analysis, we assume that the UBO density tracks the stellar distribution of each galaxy. For the Milky Way, we use the exponential Koshimoto parametric model from \citet{koshimoto_parametric_2021}. For M31, we use the stellar mass map from \citet{tamm_stellar_2012}. Due to the rapid decrease in UBO density beyond the scale height of the MW disk, the majority of UBO lenses in the MW lie in the local neighborhood of the Sun and have comparable mean bulk velocities. It is therefore the local velocity dispersion that governs the typical duration of a lensing event. We therefore take the typical velocity of UBOs in the Milky Way to be on the order of the stellar dispersion in the MW disk, $\sim 30~\rm{km/s}$ \citep{koshimoto_parametric_2021}. For the M31 velocity distribution, we note that Subaru did not take data for fields of view containing the central regions of M31 as the stellar density was too high to be able to resolve magnifications due to the lensing of a single star. Therefore, we adopt the stellar dispersion of the M31 disk, $\sim 60~\rm{km/s}$ \citep{novati_candidate_2009}, as a typical velocity for UBO lenses within M31. For the detection efficiency, we take the simulated values from Fig. 18 of the Subaru HSC Survey \citep{niikura_microlensing_2019} and perform a weighted average over the distribution of stellar magnitudes observed in patch-D2 (Fig. 22 of \citet{niikura_microlensing_2019}). This results in a typical efficiency of $\approx 0.7$ for events with $t_{\rm{FWHM}}$ in the range $0.07$ to $3$ hours. 

\begin{figure}
	\centering
	\includegraphics[width=\columnwidth]
 {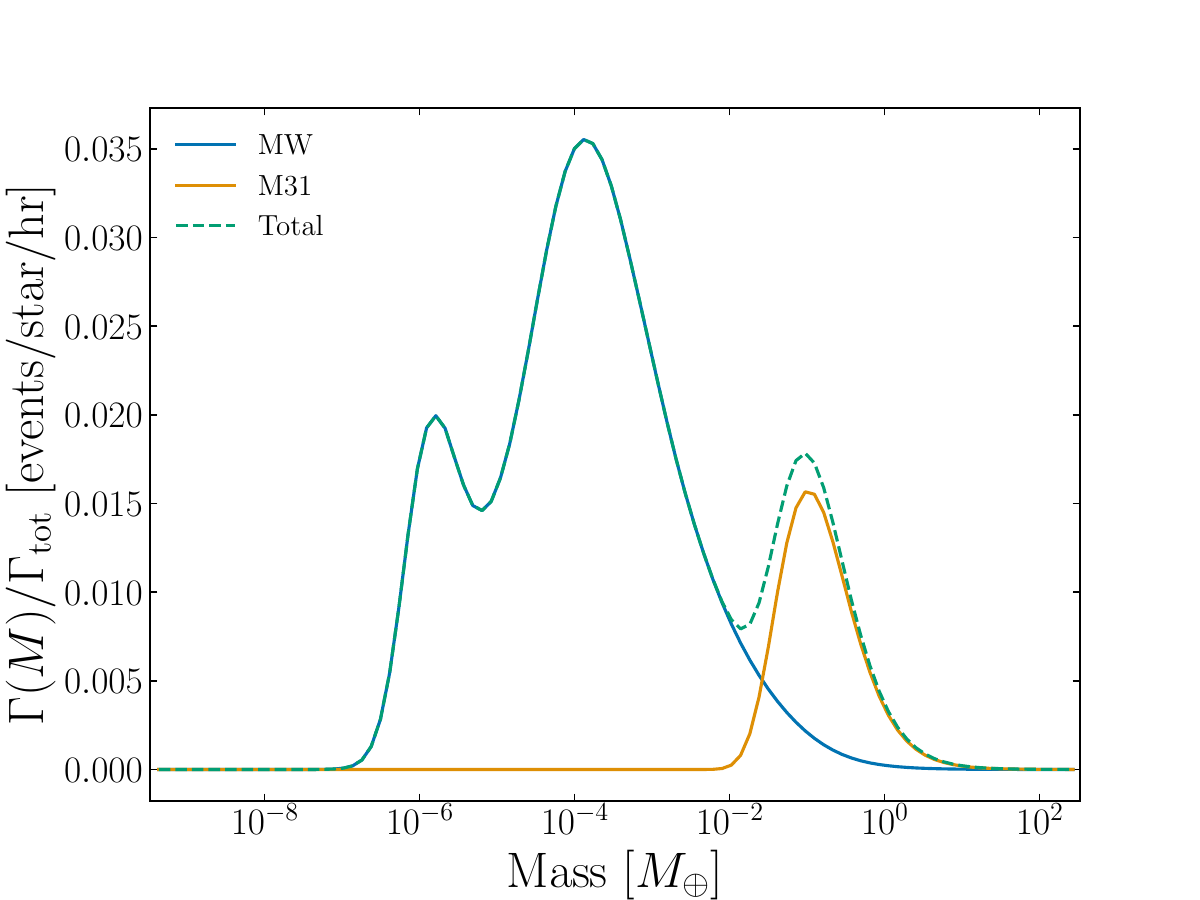}
	\caption{Expected rate of events per mass dex, normalized to total rate of events, showing the relative probability of an observed event belonging to a certain mass window. The contribution from Milky Way lenses is shown in blue and from M31 lenses in yellow, with the sum of the two shown as a green dashed curve. It is a coincidence of the survey parameters that the contributions from the Milky Way and M31 are comparable in overall probability, however they correspond to significantly different mass ranges, as discussed in appendix \ref{app:scalings}. The MW sensitivity peaks at $\mathcal{O}(10^{-4})~M_\oplus$, several orders of magnitude below the peak sensitivity of existing microlensing surveys. This is due to the more rapid observational cadence of the Subaru survey, as well as differences in the lens velocity dispersion (see appendix \ref{app:scalings}). The secondary MW peak at even lower masses arises due to the narrow region of phase space in which finite-source effects increase the event rate before ultimately suppressing the magnification below a detectable threshold.} 
	\label{fig:sensitivity}
\end{figure}

Given these assumptions, the expected differential event rate integrated over 1 dex mass bins can be seen in Fig.~\ref{fig:sensitivity}. The curves have been normalized to the total number of expected events in order to show their behavior. The most probable mass range for an observed MW lens lies within 1 dex of $10^{-4}~M_\oplus$, while the corresponding peak lies at $10^{-1}~M_\oplus$ for lenses in M31. Both of these peaks, as well as the scaling behavior at high and low masses, can be estimated analytically, as discussed in appendix~\ref{app:scalings}. The sensitivity to masses well below existing microlensing surveys is due to the high cadence of the Subaru observations, the low dispersion of MW disk lenses in comparison to the Bulge, and the large distance to M31 sources. All three factors conspire to provide peak sensitivity for the Subaru observations at masses inaccessible to existing surveys (see appendix \ref{app:scalings}).

\begin{figure}
	\centering
	\includegraphics[width=\columnwidth]
 {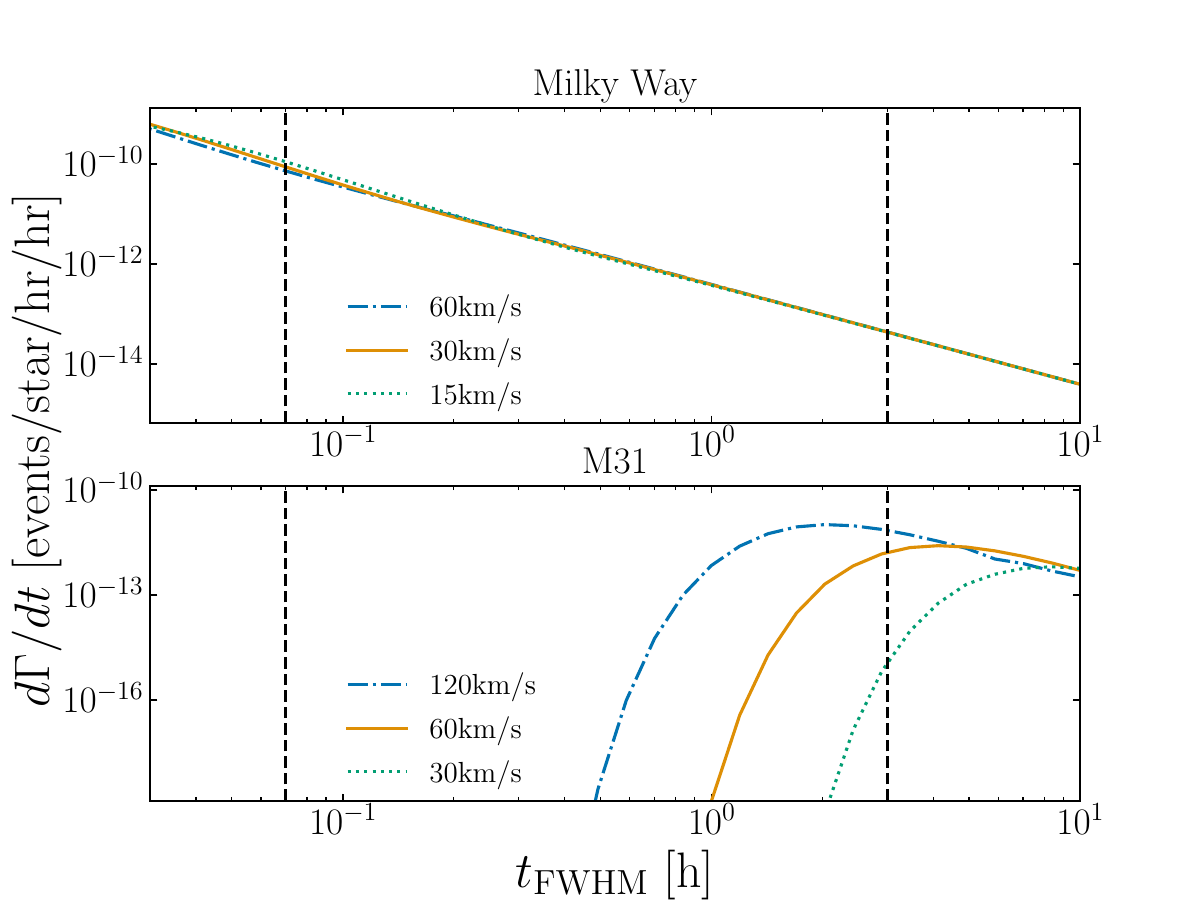}
	\caption{The effect of altering the velocity dispersion in the MW (top) and M31 (bottom) on the expected differential lensing event rate for our fiducial $p = 1$ mass distribution (Eq.~\ref{eq:fid}). The effect is minimal in the MW due to the $\propto v_T^{-2(p-1)}$ scaling of the rate. However, the low-mass cutoff in M31 induced by finite-source effects leads to large changes in the distribution under varying dispersions. See section~\ref{sec:expected_event_rate} for details.}
	\label{fig:differential_rate_velocity}
\end{figure}

While we have chosen particular models for the UBO spatial and velocity distributions in each galaxy in order to compute our limits, our results are largely insensitive to the specifics of the chosen model. For the MW, the uncertainty in stellar density from the Koshimoto model is encapsulated in the total mid-plane stellar density of $0.040 \pm 0.002 ~M_{\odot} ~\rm{pc}^{-3}$, a difference of 5\% in either direction \citep{bovy_stellar_2017}. For M31, the densities used are lower limits with the upper limits being a factor of $1.5$ times higher \citep{tamm_stellar_2012}.  
Varying the velocity dispersion leads to a larger effect, as can be seen in Fig.~\ref{fig:differential_rate_velocity}, which plots the distribution of $t_{\rm{FWHM}}$ for both the MW (top) and M31 (bottom). In the top panel, changing the dispersion does not induce an apparent change in the distribution of $t_{\rm{FWHM}}$, as the lensing rate $d\Gamma/d t \propto v_T^{-2(p-1)}$,\footnote{The lensing rate per star $\Gamma$ at a given distance $d$ is proportional $N$ (the number of lenses in the plane), $R_E$ (the ``cross-section'' for a lens to cross the star in the lensing plane), and $v_T$ (the speed with which the crossing takes place). Taking $d/dM$ yields $\frac{d\Gamma}{dM} \propto \frac{dN}{dM} R_E v_T$. With $R_E \propto M^{1/2}$ (eq.~\ref{eq:RE}), $\frac{dN}{dM} \propto M^{-p-1}$ (eq.~\ref{eq:mass_func}), and $M\propto t^2 v_T^2$ (eq.~\ref{eq:tE}), one recovers $\frac{d\Gamma}{dt} \propto t^{-2p}v_T^{-2(p-1)}$.} hence for our $p=1$ fiducial power-law (eq.~\ref{eq:fid}), the $v_T$ scaling drops out. In M31 (bottom), however, finite-source effects set a scale, limiting sensitivity at low masses. Thus, varying the velocity dispersion does change the distribution even for $p=1$, with higher dispersions shifting the distribution to smaller timescales. The vertical black dashed lines correspond to the range of $t_{\rm{FWHM}}$ to which the survey is sensitive, indicating the loss in sensitivity for high dispersions. Empirically, changing the dispersion in the MW and M31 by a factor of two in either direction leads to a difference of $\sim 40 \%$ in the total event rate compared to our fiducial values. 

\section{Results}
\label{sec:results}

We place limits in a manner akin to \citet{niikura_microlensing_2019}. We begin by computing the expected event rate for the HSC observation using equation \ref{eq:difRate} for a power-law probability distribution $f(M) = A \Big( \frac{M}{M_{\oplus}}\Big)^{-p}$ for $M \in [M_{\rm{min}}, M_{\rm{max}}]$, with the prefactor 
\begin{equation}
    A = \Big(\int_{M_{\rm{min}}}^{M_{\rm{max}}} \Big( \frac{M}{M_{\oplus}}\Big)^{-p}\Big)^{-1}
\end{equation}
ensuring normalization. We take $M_{\rm{min}} = 3.3 \times 10^{-10} ~M_{\oplus},~ M_{\rm{max}} = 3.3 \times 10^{2} ~M_{\oplus}$. Note that while for computational reasons we truncate the distribution, our results are insensitive to these bounds, as these are well outside our mass-range of peak sensitivity, $10^{-2}~M_\oplus \lesssim M \lesssim M_\oplus$ (see Fig.~\ref{fig:sensitivity}).

\begin{figure}
	\centering
	\includegraphics[width=\columnwidth]
 {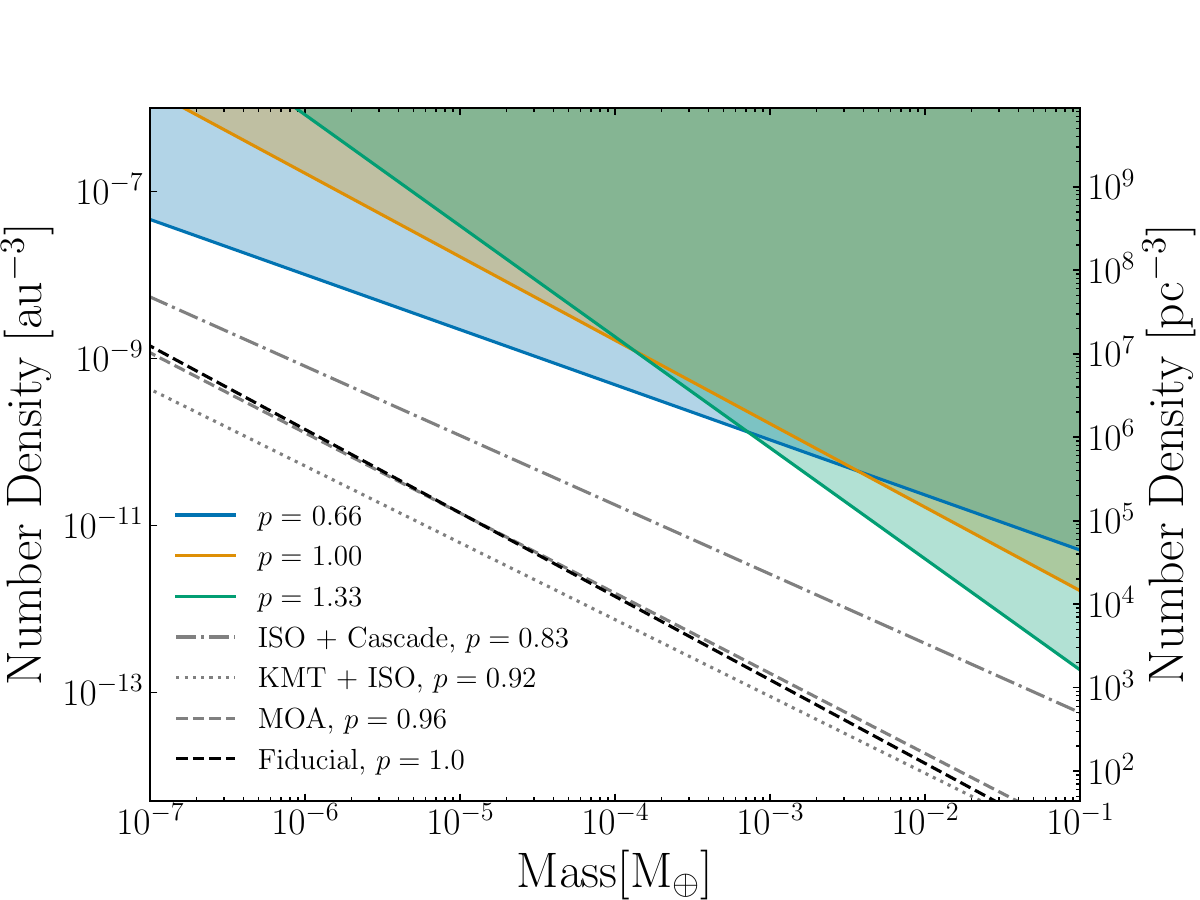}
	\caption{Limits on the local number density of UBOs within 1 dex of $M$. The solid lines correspond to the upper limit for three different values of the power-law index $p$ of the UBO mass function placed by the Subaru HSC observations. The gray and black lines correspond to the mass functions enumerated in section~\ref{sec:abundance}. Note that the \texttt{ISO+cascade} results are derived from asteroid-mass objects while the KMTNet and MOA results are derived using data at masses $M > M_\oplus$, hence all are extrapolations in the mass range probed by Subaru.}
	\label{fig:n_limits_mass}
\end{figure}

By integrating the differential rate, equation \ref{eq:difRate}, over the 7-hour observational time range, we calculate the overall expected number of events. Microlensing events follow a Poisson distribution with an estimated number of observations given by 
\begin{equation}
    P(k = N_{\rm{obs}}|N_{\rm{exp}}) = \frac{N_{\rm{exp}}^k}{k!}e^{-N_{\rm{exp}}}.
\end{equation} 
The survey produced a single candidate event. This leads to a  95\% confidence interval given by $P(k=0) + P(k=1) \geq 0.05$, which corresponds to $N_{\rm{exp}} < 4.74$ \citep{niikura_microlensing_2019}. This criterion can then be used for to place a constraint on the abundance of high-mass UBOs for a particular value of $p$, the power-law index of the UBO mass function. These constraints place a robust upper limit on the abundance of UBOs in the mass range of $10^{-5}~M_\oplus - 1~M_\oplus$ kg. We plot the results in Fig.~\ref{fig:n_limits_mass}. The solid lines correspond to three different values of the power-law index $p$ of the UBO mass function in the range $0.66 < p < 1.33$. The shaded regions above each curve are excluded by our analysis. The limit on UBOs with mass within 1 dex of $10^{-4} ~M_{\oplus}$ is $ < 1.4 \times 10^{7} ~\rm{pc}^{-3}$. The dash-dotted gray line is the mass function of UBOs when adopting the Do et al. normalization derived from the observation of 1I/`Oumuamua and extrapolating to high masses with the theoretical prediction for a collisional cascade $p = 0.83$ (Eq.~\ref{eq:dohnanyi}). The gray dashed and dotted lines are the best-fit mass functions derived by MOA and KMTNet respectively using data at masses $M > M_\oplus$ (Eqs.~\ref{eq:moa} and \ref{eq:kmt} respectively). The black dashed line is our fiducial mass function (Eq.~\ref{eq:fid}). Note that, as discussed in section~\ref{sec:massdist}, all three of these curves are extrapolations in the sub-terrestrial range. Our limits have support in this range, hence provide a direct probe of the UBO abundance at these masses.

\begin{figure}
	\centering
	\includegraphics[width=\columnwidth]
 {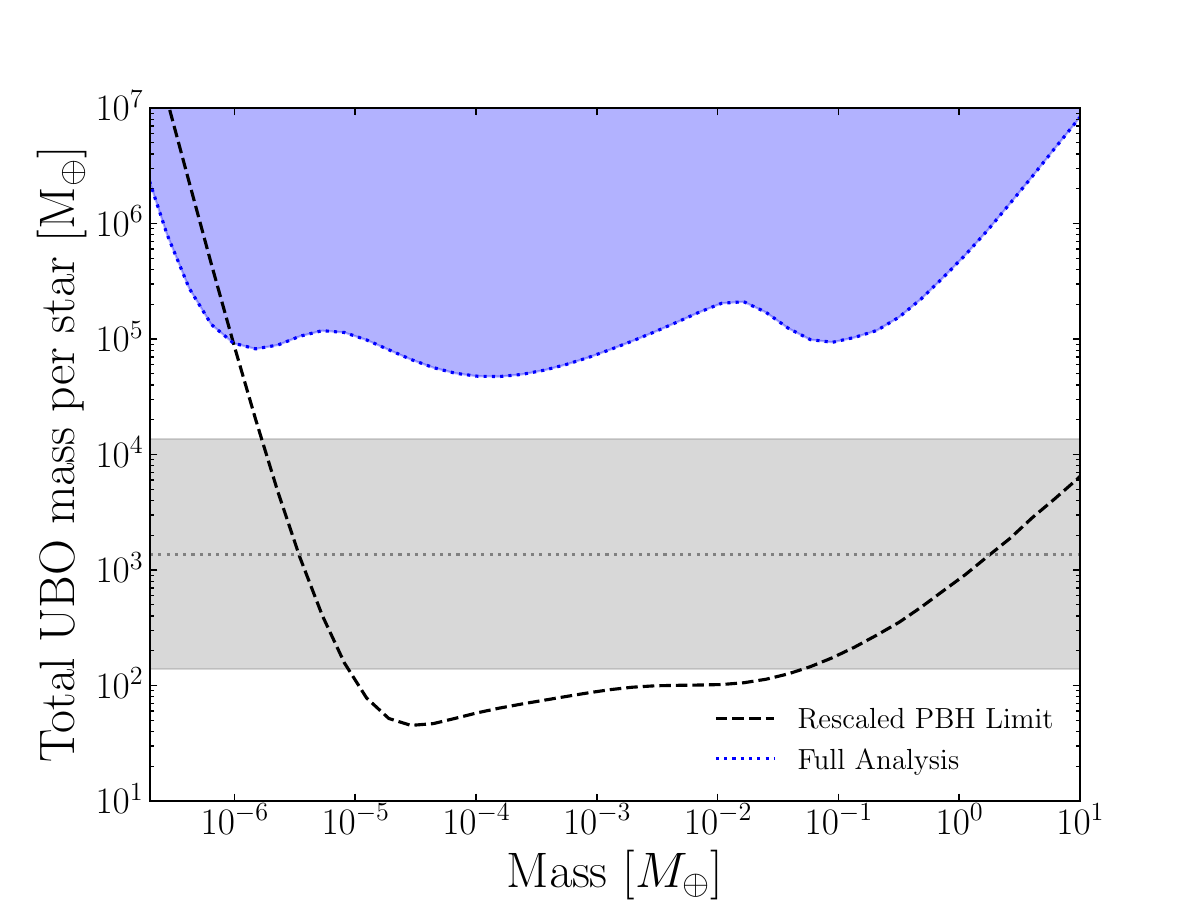}
	\caption{ Limits on the total abundance of UBOs from the Subaru HSC survey for a monochromatic mass spectrum. The curves correspond to the limit on ejected mass per star at masses within 1 dex of $M$. The black dashed line shows a previous limit derived by rescaling the existing Subaru constraints on primordial black holes, which implicitly assumed the UBO lenses follow the halo-like distribution of dark matter. Our corrected constraints (blue) are derived using the expected spatial density of UBOs. See Fig.~\ref{fig:line_of_sight} for a comparison of the UBO and dark matter line-of-sight densities.  The gray band corresponds to typical protoplanetary disk masses and serves as a point of reference for theoretically-motivated regions of UBO density (see main text).}
	\label{fig:monochromatic_mass}
\end{figure}

We also show the results of our analysis in the ``monochromatic limit'' in Fig.~\ref{fig:monochromatic_mass}. At a given mass $M$, we assume the number density of UBOs to be approximately constant 1 dex of $M$ and zero elsewhere. We then apply the condition $N_{\rm{exp}} < 4.74$ to this truncated distribution to set a ``monochromatic'' limit. The resulting limits are shown as a function of mass in Fig.~\ref{fig:monochromatic_mass} in blue. While a narrow distribution of UBO masses is not physically-motivated, we choose to display this result to compare to existing literature (shown as the black dashed line) in which the Subaru PBH constraints have been rescaled to UBOs without consideration of the differences in the spatial distributions of the two lens populations. The limit computed in this work, taking into account differences in the lens populations, is clearly significantly weaker than the naive estimate. This is due in large part to the low relative density of UBOs along the MW--M31 line-of-sight with respect to the PBH density. The weakened sensitivity at $10^{-2}~M_\oplus$ is due to the gap between the distributions of events from MW and M31 lenses seen in Fig. \ref{fig:sensitivity}.

We wish to note that while the Subaru observations we used for our analysis provide a direct observational probe of UBOs in the sub-terrestrial range, due to the short baseline and low UBO density along the line of sight, they are not ideal for constraining the UBO abundance. As such, the resulting limits do not probe abundances motivated by our current understanding of planetary formation.
This is evident in Fig.~\ref{fig:monochromatic_mass}, where we plot a gray band corresponding to an estimate of the average initial protoplanetary disk mass of stars in the local neighborhood \citep{Williams_2011}, using an approximate average stellar mass of $0.5~M_\odot$ \citep{Golovin_2023,bovy_stellar_2017}. Since UBOs are thought to be comprised of material from these disks, the total UBO mass per star is expected to lie within or below the gray band, while our limits lie one to two orders of magnitude above.

While our results do not at present place strong constraints on models of UBO formation, they demonstrate that even small differences in cadence can give rise to significant improvements in sensitivity to the UBO mass function, motivating the use of the most rapid achievable cadence in future surveys. Additionally, they show that the microlensing event observed by \citet{niikura_microlensing_2019} was very unlikely to have been caused by a UBO and motivate a careful reappraisal of the event, as it may be a hint of a more exotic source.

\section{Discussion and conclusions}
\label{sec:discussion}

The abundance of unbound objects in the sub-terrestrial mass range provides a window into the dynamical processes occurring during planetary formation. However, due to their low mass, such UBOs are typically unobservable for existing microlensing surveys due to their $> 15$-minute minimum observational cadence. In this paper, we have leveraged existing observations performed by the Subaru telescope at 2-minute cadence to constrain the abundance of unbound objects with masses in the range $10^{-5}~M_\oplus - 1~M_\oplus$ accounting, for the first time, for the UBO spatial distribution. Furthermore, we have demonstrated that a faster cadence by one order of magnitude can provide a several orders-of-magnitude detection boost for low-mass UBOs. These results motivate future ultra-high-cadence surveys to better characterize the sub-terrestrial UBO population. Implementing such an observational strategy in upcoming surveys has the potential to yield observations of unbound planetesimals deep in the sub-terrestrial regime.

\section*{Acknowledgements}

The authors would like to thank the anonymous referee for their useful commentary on early versions of this manuscript, which helped to improve its quality. This material is based upon work supported in part by the National Science Foundation Graduate Research Fellowship under Grant No. DGE-1842400 (NS) and the U.S. Department of Energy grant number de-sc0010107 (WD and SP).

%%%%%%%%%%%%%%%%%%%%%%%%%%%%%%%%%%%%%%%%%%%%%%%%%%
\section*{Data Availability}

All code used in this analysis has been made publicly available on GitHub.\footnote{\texttt{https://github.com/NolanSmyth/LensCalcPy}}


\bibliographystyle{mnras}
\bibliography{main}

\begin{thebibliography}{}
\makeatletter
\relax
\def\mn@urlcharsother{\let\do\@makeother \do\$\do\&\do\#\do\^\do\_\do\%\do\~}
\def\mn@doi{\begingroup\mn@urlcharsother \@ifnextchar [ {\mn@doi@}
  {\mn@doi@[]}}
\def\mn@doi@[#1]#2{\def\@tempa{#1}\ifx\@tempa\@empty \href
  {http://dx.doi.org/#2} {doi:#2}\else \href {http://dx.doi.org/#2} {#1}\fi
  \endgroup}
\def\mn@eprint#1#2{\mn@eprint@#1:#2::\@nil}
\def\mn@eprint@arXiv#1{\href {http://arxiv.org/abs/#1} {{\tt arXiv:#1}}}
\def\mn@eprint@dblp#1{\href {http://dblp.uni-trier.de/rec/bibtex/#1.xml}
  {dblp:#1}}
\def\mn@eprint@#1:#2:#3:#4\@nil{\def\@tempa {#1}\def\@tempb {#2}\def\@tempc
  {#3}\ifx \@tempc \@empty \let \@tempc \@tempb \let \@tempb \@tempa \fi \ifx
  \@tempb \@empty \def\@tempb {arXiv}\fi \@ifundefined
  {mn@eprint@\@tempb}{\@tempb:\@tempc}{\expandafter \expandafter \csname
  mn@eprint@\@tempb\endcsname \expandafter{\@tempc}}}

\bibitem[\protect\citeauthoryear{Abe et~al.}{Abe et~al.}{2008}]{Abe:2008rpq}
Abe F.,  et~al., 2008, in {17th Workshop on General Relativity and Gravitation
  in Japan}. pp 62--74

\bibitem[\protect\citeauthoryear{Bhatiani, Dai  \& Guerras}{Bhatiani
  et~al.}{2019}]{Bhatiani_2019}
Bhatiani S.,  Dai X.,   Guerras E.,  2019, \mn@doi [The Astrophysical Journal]
  {10.3847/1538-4357/ab46ac}, 885, 77

\bibitem[\protect\citeauthoryear{Bland-Hawthorn \& Gerhard}{Bland-Hawthorn \&
  Gerhard}{2016}]{doi:10.1146/annurev-astro-081915-023441}
Bland-Hawthorn J.,  Gerhard O.,  2016, \mn@doi [Annual Review of Astronomy and
  Astrophysics] {10.1146/annurev-astro-081915-023441}, 54, 529

\bibitem[\protect\citeauthoryear{{Borisov} \& {Shustov}}{{Borisov} \&
  {Shustov}}{2021}]{2021SoSyR..55..124B}
{Borisov} G.~V.,  {Shustov} B.~M.,  2021, \mn@doi [Solar System Research]
  {10.1134/S0038094621020027}, \href
  {https://ui.adsabs.harvard.edu/abs/2021SoSyR..55..124B} {55, 124}

\bibitem[\protect\citeauthoryear{Bovy}{Bovy}{2017}]{bovy_stellar_2017}
Bovy J.,  2017, \mn@doi [Monthly Notices of the Royal Astronomical Society]
  {10.1093/mnras/stx1277}, 470, 1360

\bibitem[\protect\citeauthoryear{Brown \& Borovicka}{Brown \&
  Borovicka}{2023}]{brown2023proposed}
Brown P.~G.,  Borovicka J.,  2023, On the Proposed Interstellar Origin of the
  USG 20140108 Fireball (\mn@eprint {arXiv} {2306.14267})

\bibitem[\protect\citeauthoryear{Carr \& Kuhnel}{Carr \&
  Kuhnel}{2022}]{Carr_2022}
Carr B.,  Kuhnel F.,  2022, \mn@doi [{SciPost} Physics Lecture Notes]
  {10.21468/scipostphyslectnotes.48}

\bibitem[\protect\citeauthoryear{Croon, McKeen, Raj  \& Wang}{Croon
  et~al.}{2020}]{croon_subaru_2020}
Croon D.,  McKeen D.,  Raj N.,   Wang Z.,  2020, \mn@doi [Physical Review D]
  {10.1103/PhysRevD.102.083021}, 102, 083021

\bibitem[\protect\citeauthoryear{Dai \& Guerras}{Dai \&
  Guerras}{2018}]{Dai_2018}
Dai X.,  Guerras E.,  2018, \mn@doi [The Astrophysical Journal]
  {10.3847/2041-8213/aaa5fb}, 853, L27

\bibitem[\protect\citeauthoryear{Do, Tucker  \& Tonry}{Do
  et~al.}{2018}]{Do_2018}
Do A.,  Tucker M.~A.,   Tonry J.,  2018, \mn@doi [The Astrophysical Journal]
  {10.3847/2041-8213/aaae67}, 855, L10

\bibitem[\protect\citeauthoryear{{Dohnanyi}}{{Dohnanyi}}{1968}]{1968IAUS...33..486D}
{Dohnanyi} J.~S.,  1968, in {Kresak} L.,  {Millman} P.~M.,  eds,  Vol. 33,
  Physics and Dynamics of Meteors. p.~486

\bibitem[\protect\citeauthoryear{Engelhardt, Jedicke, Vere{\v{s} },
  Fitzsimmons, Denneau, Beshore  \& Meinke}{Engelhardt
  et~al.}{2017}]{Engelhardt_2017}
Engelhardt T.,  Jedicke R.,  Vere{\v{s} } P.,  Fitzsimmons A.,  Denneau L.,
  Beshore E.,   Meinke B.,  2017, \mn@doi [The Astronomical Journal]
  {10.3847/1538-3881/aa5c8a}, 153, 133

\bibitem[\protect\citeauthoryear{Fitzsimmons, Meech, Matra  \&
  Pfalzner}{Fitzsimmons et~al.}{2023}]{fitzsimmons2023interstellar}
Fitzsimmons A.,  Meech K.,  Matra L.,   Pfalzner S.,  2023, Interstellar
  Objects and Exocomets (\mn@eprint {arXiv} {2303.17980})

\bibitem[\protect\citeauthoryear{G{\'{a}}sp{\'{a}}r, Psaltis, Rieke  \&
  Ozel}{G{\'{a}}sp{\'{a}}r et~al.}{2012}]{G_sp_r_2012}
G{\'{a}}sp{\'{a}}r A.,  Psaltis D.,  Rieke G.~H.,   Ozel F.,  2012, \mn@doi
  [The Astrophysical Journal] {10.1088/0004-637x/754/1/74}, 754, 74

\bibitem[\protect\citeauthoryear{Golovin, Reffert, Just, Jordan, Vani  \&
  Jahrei{\ss}}{Golovin et~al.}{2023}]{Golovin_2023}
Golovin A.,  Reffert S.,  Just A.,  Jordan S.,  Vani A.,   Jahrei{\ss} H.,
  2023, \mn@doi [Astronomy {\&} Astrophysics] {10.1051/0004-6361/202244250},
  670, A19

\bibitem[\protect\citeauthoryear{Gould et~al.,}{Gould
  et~al.}{2022}]{10.5303/JKAS.2022.55.5.173}
Gould A.,  et~al., 2022, Journal of The Korean Astronomical Society, 55, 173

\bibitem[\protect\citeauthoryear{Hui, Ye, F{\"o}hring, Hung  \& Tholen}{Hui
  et~al.}{2020}]{Hui_2020}
Hui M.-T.,  Ye Q.-Z.,  F{\"o}hring D.,  Hung D.,   Tholen D.~J.,  2020, \mn@doi
  [The Astronomical Journal] {10.3847/1538-3881/ab9df8}, 160, 92

\bibitem[\protect\citeauthoryear{Jennings, Cordes  \& Chatterjee}{Jennings
  et~al.}{2020}]{Jennings_2020}
Jennings R.~J.,  Cordes J.~M.,   Chatterjee S.,  2020, \mn@doi [The
  Astrophysical Journal] {10.3847/1538-4357/ab64df}, 889, 145

\bibitem[\protect\citeauthoryear{Jewitt, Luu, Rajagopal, Kotulla, Ridgway, Liu
  \& Augusteijn}{Jewitt et~al.}{2017}]{Jewitt_2017}
Jewitt D.,  Luu J.,  Rajagopal J.,  Kotulla R.,  Ridgway S.,  Liu W.,
  Augusteijn T.,  2017, \mn@doi [The Astrophysical Journal Letters]
  {10.3847/2041-8213/aa9b2f}, 850, L36

\bibitem[\protect\citeauthoryear{Johnson, Penny, Gaudi, Kerins, Rattenbury,
  Robin, Novati  \& Henderson}{Johnson et~al.}{2020}]{johnson_predictions_2020}
Johnson S.~A.,  Penny M.~T.,  Gaudi B.~S.,  Kerins E.,  Rattenbury N.~J.,
  Robin A.~C.,  Novati S.~C.,   Henderson C.~B.,  2020, \mn@doi [The
  Astronomical Journal] {10.3847/1538-3881/aba75b}, 160, 123

\bibitem[\protect\citeauthoryear{{Kim} et~al.,}{{Kim}
  et~al.}{2016}]{2016JKAS...49...37K}
{Kim} S.-L.,  et~al., 2016, \mn@doi [Journal of Korean Astronomical Society]
  {10.5303/JKAS.2016.49.1.37}, \href
  {https://ui.adsabs.harvard.edu/abs/2016JKAS...49...37K} {49, 37}

\bibitem[\protect\citeauthoryear{Koshimoto, Baba  \& Bennett}{Koshimoto
  et~al.}{2021}]{koshimoto_parametric_2021}
Koshimoto N.,  Baba J.,   Bennett D.~P.,  2021, \mn@doi [The Astrophysical
  Journal] {10.3847/1538-4357/ac07a8}, 917, 78

\bibitem[\protect\citeauthoryear{Koshimoto et~al.,}{Koshimoto
  et~al.}{2023}]{koshimoto2023terrestrial}
Koshimoto N.,  et~al., 2023, Terrestrial and Neptune mass free-floating planet
  candidates from the MOA-II 9-year Galactic Bulge survey (\mn@eprint {arXiv}
  {2303.08279})

\bibitem[\protect\citeauthoryear{Landgraf, Baggaley, Grun, Kruger  \&
  Linkert}{Landgraf et~al.}{2000}]{Landgraf_2000}
Landgraf M.,  Baggaley W.~J.,  Grun E.,  Kruger H.,   Linkert G.,  2000,
  \mn@doi [Journal of Geophysical Research: Space Physics]
  {10.1029/1999ja900359}, 105, 10343

\bibitem[\protect\citeauthoryear{Lohne, Krivov  \& Rodmann}{Lohne
  et~al.}{2008}]{Lohne_2008}
Lohne T.,  Krivov A.~V.,   Rodmann J.,  2008, \mn@doi [The Astrophysical
  Journal] {10.1086/524840}, 673, 1123

\bibitem[\protect\citeauthoryear{Matsunaga \& Yamamoto}{Matsunaga \&
  Yamamoto}{2006}]{matsunaga_finite_2006}
Matsunaga N.,  Yamamoto K.,  2006, \mn@doi [Journal of Cosmology and
  Astroparticle Physics] {10.1088/1475-7516/2006/01/023}, 2006, 023

\bibitem[\protect\citeauthoryear{{Meech} et~al.,}{{Meech}
  et~al.}{2017}]{2017Natur.552..378M}
{Meech} K.~J.,  et~al., 2017, \mn@doi [\nat] {10.1038/nature25020}, \href
  {https://ui.adsabs.harvard.edu/abs/2017Natur.552..378M} {552, 378}

\bibitem[\protect\citeauthoryear{{Moro-Mart{\'\i}n}, {Turner}  \&
  {Loeb}}{{Moro-Mart{\'\i}n} et~al.}{2009}]{2009ApJ...704..733M}
{Moro-Mart{\'\i}n} A.,  {Turner} E.~L.,   {Loeb} A.,  2009, \mn@doi [\apj]
  {10.1088/0004-637X/704/1/733}, \href
  {https://ui.adsabs.harvard.edu/abs/2009ApJ...704..733M} {704, 733}

\bibitem[\protect\citeauthoryear{{Mr{\'{o}}z} et~al.,}{{Mr{\'{o}}z}
  et~al.}{2019}]{2019A&A...622A.201M}
{Mr{\'{o}}z} P.,  et~al., 2019, \mn@doi [\aap] {10.1051/0004-6361/201834557},
  \href {https://ui.adsabs.harvard.edu/abs/2019A&A...622A.201M} {622, A201}

\bibitem[\protect\citeauthoryear{Mr{\'{o}}z et~al.,}{Mr{\'{o}}z
  et~al.}{2020}]{Mroz_2020}
Mr{\'{o}}z P.,  et~al., 2020, \mn@doi [The Astrophysical Journal Letters]
  {10.3847/2041-8213/abbfad}, 903, L11

\bibitem[\protect\citeauthoryear{Nakamura \& Deguchi}{Nakamura \&
  Deguchi}{1999}]{nakamura_wave_1999}
Nakamura T.~T.,  Deguchi S.,  1999, \mn@doi [Progress of Theoretical Physics
  Supplement] {10.1143/PTPS.133.137}, 133, 137

\bibitem[\protect\citeauthoryear{Niikura et~al.,}{Niikura
  et~al.}{2019}]{niikura_microlensing_2019}
Niikura H.,  et~al., 2019, \mn@doi [Nature Astronomy]
  {10.1038/s41550-019-0723-1}, 3, 524

\bibitem[\protect\citeauthoryear{Novati}{Novati}{2009}]{Calchi_Novati_2009}
Novati S.~C.,  2009, \mn@doi [General Relativity and Gravitation]
  {10.1007/s10714-009-0918-3}, 42, 2101

\bibitem[\protect\citeauthoryear{Novati et~al.,}{Novati
  et~al.}{2009}]{novati_candidate_2009}
Novati S.~C.,  et~al., 2009, \mn@doi [The Astrophysical Journal]
  {10.1088/0004-637X/695/1/442}, 695, 442

\bibitem[\protect\citeauthoryear{Paczynski}{Paczynski}{1986}]{paczynski_gravitational_1986}
Paczynski B.,  1986, \mn@doi [ApJ] {10.1086/163919}, 301, 503

\bibitem[\protect\citeauthoryear{Seligman \& Laughlin}{Seligman \&
  Laughlin}{2020}]{Seligman_2020}
Seligman D.,  Laughlin G.,  2020, \mn@doi [The Astrophysical Journal Letters]
  {10.3847/2041-8213/ab963f}, 896, L8

\bibitem[\protect\citeauthoryear{{Siraj} \& {Loeb}}{{Siraj} \&
  {Loeb}}{2019}]{2019arXiv190603270S}
{Siraj} A.,  {Loeb} A.,  2019, \mn@doi [arXiv e-prints]
  {10.48550/arXiv.1906.03270}, \href
  {https://ui.adsabs.harvard.edu/abs/2019arXiv190603270S} {p. arXiv:1906.03270}

\bibitem[\protect\citeauthoryear{Siraj \& Loeb}{Siraj \&
  Loeb}{2022}]{siraj20222019}
Siraj A.,  Loeb A.,  2022, The 2019 Discovery of a Meteor of Interstellar
  Origin (\mn@eprint {arXiv} {1904.07224})

\bibitem[\protect\citeauthoryear{Smyth, Profumo, English, Jeltema, McKinnon  \&
  Guhathakurta}{Smyth et~al.}{2020}]{smyth_updated_2020}
Smyth N.,  Profumo S.,  English S.,  Jeltema T.,  McKinnon K.,   Guhathakurta
  P.,  2020, \mn@doi [Physical Review D] {10.1103/PhysRevD.101.063005}, 101,
  063005

\bibitem[\protect\citeauthoryear{{Sterken}, {Westphal}, {Altobelli},
  {Malaspina}  \& {Postberg}}{{Sterken} et~al.}{2019}]{2019SSRv..215...43S}
{Sterken} V.~J.,  {Westphal} A.~J.,  {Altobelli} N.,  {Malaspina} D.,
  {Postberg} F.,  2019, \mn@doi [\ssr] {10.1007/s11214-019-0607-9}, \href
  {https://ui.adsabs.harvard.edu/abs/2019SSRv..215...43S} {215, 43}

\bibitem[\protect\citeauthoryear{Strigari, Barnab{\`{e} }, Marshall  \&
  Blandford}{Strigari et~al.}{2012}]{Strigari_2012}
Strigari L.~E.,  Barnab{\`{e} } M.,  Marshall P.~J.,   Blandford R.~D.,  2012,
  \mn@doi [Monthly Notices of the Royal Astronomical Society]
  {10.1111/j.1365-2966.2012.21009.x}, 423, 1856

\bibitem[\protect\citeauthoryear{Sugiyama, Kurita  \& Takada}{Sugiyama
  et~al.}{2019}]{sugiyama_revisiting_2019}
Sugiyama S.,  Kurita T.,   Takada M.,  2019, arXiv:1905.06066 [astro-ph]

\bibitem[\protect\citeauthoryear{Sumi et~al.,}{Sumi
  et~al.}{2023}]{sumi2023freefloating}
Sumi T.,  et~al., 2023, Free-Floating planet Mass Function from MOA-II 9-year
  survey towards the Galactic Bulge (\mn@eprint {arXiv} {2303.08280})

\bibitem[\protect\citeauthoryear{Tamm, Tempel, Tenjes, Tihhonova  \&
  Tuvikene}{Tamm et~al.}{2012}]{tamm_stellar_2012}
Tamm A.,  Tempel E.,  Tenjes P.,  Tihhonova O.,   Tuvikene T.,  2012, \mn@doi
  [Astronomy \& Astrophysics] {10.1051/0004-6361/201220065}, 546, A4

\bibitem[\protect\citeauthoryear{Trilling et~al.,}{Trilling
  et~al.}{2017}]{Trilling_2017}
Trilling D.~E.,  et~al., 2017, \mn@doi [The Astrophysical Journal]
  {10.3847/2041-8213/aa9989}, 850, L38

\bibitem[\protect\citeauthoryear{Udalski, Szyma{\'{n}}ski  \&
  Szyma{\'{n}}ski}{Udalski et~al.}{2015}]{udalski2015ogleiv}
Udalski A.,  Szyma{\'{n}}ski M.~K.,   Szyma{\'{n}}ski G.,  2015, OGLE-IV:
  Fourth Phase of the Optical Gravitational Lensing Experiment (\mn@eprint
  {arXiv} {1504.05966})

\bibitem[\protect\citeauthoryear{{Vaubaillon}}{{Vaubaillon}}{2022}]{2022JIMO...50..140V}
{Vaubaillon} J.,  2022, \mn@doi [WGN, Journal of the International Meteor
  Organization] {10.48550/arXiv.2211.02305}, \href
  {https://ui.adsabs.harvard.edu/abs/2022JIMO...50..140V} {50, 140}

\bibitem[\protect\citeauthoryear{Williams \& Cieza}{Williams \&
  Cieza}{2011}]{Williams_2011}
Williams J.~P.,  Cieza L.~A.,  2011, \mn@doi [Annual Review of Astronomy and
  Astrophysics] {10.1146/annurev-astro-081710-102548}, 49, 67

\bibitem[\protect\citeauthoryear{Witt \& Mao}{Witt \&
  Mao}{1994}]{witt_can_1994}
Witt H.~J.,  Mao S.,  1994, \mn@doi [The Astrophysical Journal]
  {10.1086/174426}, 430, 505

\makeatother
\end{thebibliography}

%%%%%%%%%%%%%%%%%%%%%%%%%%%%%%%%%%%%%%%%%%%%%%%%%%

%%%%%%%%%%%%%%%%% APPENDICES %%%%%%%%%%%%%%%%%%%%%

\appendix

\section{Differential event rate scaling}
\label{app:scalings}

The probability density of an observed event corresponding to a particular mass $M$, marginalizing over all other parameters, is proportional to the lensing rate $\Gamma$. The differential lensing rate is given by 
\begin{multline}
    \label{eq:difRateAll}
     \frac{d\Gamma}{dM\,dd_L\,dt_{\rm{FWHM}}\,du_0} =  
     \frac{2}{\sqrt{u_T^2 - u_0^2}} \frac{v_T^4}{v_c^2} \exp \Big[ -\frac{v_T^2}{v_c^2}\Big] \frac{\rho_M}{M}f(M) \varepsilon(t_{\rm{FWHM}}),
\end{multline}
where $f(M)$ is the probability distribution function of lens masses, $t_{\rm{FWHM}}$ is the full-width-half-maximum duration, $u_0$ is the impact parameter at the point of closest approach between the lens and source, $u_T$ is the largest impact parameter for which the magnification remains above a detectable threshold ($A > 1.34$ in our analysis), $\rho_M$ is the mass density of the lens population, $d_S$ is the distance to the source star, and $\varepsilon(t_{\rm{FWHM}})$ is the detection efficiency. The transverse velocity of the lens, $v_T$, is given by
\begin{equation}
\label{eq:vT}
 v_T = 2R_E \sqrt{u_{\rm{HM}}^2 - u_{\min}^2}/t_{\rm{FWHM}}.
\end{equation}
To derive the peak sensitivity with respect to mass, we marginalize the differential lensing rate over $d_L, t_{\rm{FWHM}},$ and $u_0$:
\begin{multline}
    \label{eq:difRate_mass}
     \frac{d\Gamma}{dM}(M) = 2 \int_0^{d_s}dd_L \int_0^{u_T}du_0 \int_{t_\text{min}}^{t_\text{max}}dt_{\rm{FWHM}}\\
     \frac{1}{\sqrt{u_T^2 - u_0^2}} \frac{v_T^4}{v_c^2} \exp \Big[ -\frac{v_T^2}{v_c^2}\Big] \frac{\rho_M}{M}f(M) \varepsilon(t_{\rm{FWHM}}).
\end{multline}
Here,
$t_\text{min}$ is the minimum detectable timescale for an event and is usually roughly $t_\text{cad}$, the observational cadence. The longest timescale of a detectable event, $t_\text{max}$, is set to the total duration of the survey, $t_\text{obs}$. Note that for events with $t_{\rm{FWHM}} \approx t_\text{obs}$, the efficiency decreases, as the event may extend beyond the observational window.

Figure \ref{fig:diff_rate_M_MW} shows $\frac{d\Gamma}{dM}$ as a function of mass $M$ for lenses within the Milky Way. The actual curve for the Subaru observations is shown in solid black. We also display in  dashed blue the equivalent curve if all finite-source effects are neglected. It is clear that finite-source effects severely  limit sensitivity at low masses, as discussed below. At high masses, the decrease in lensing rate is due to the finite cadence, as can be seen from the yellow dotted curve, which is the lensing rate for the Subaru observation if the cadence had been increased by an order of magnitude ($t_\text{min} = 60$ minutes). The green dashed curve is the lensing rate for a $t_\text{min} = 60$-minute observation neglecting finite-source effects.

The scaling of both high-mass and low-mass regimes can be estimated in a simple fashion, derived below. In the following, we work in natural units with $c=1$.

\subsection{Low-mass behavior}
\label{app:lowmass}

On the low-mass end, recall that finite-source effects become relevant when $\rho = \frac{R_S/d_S}{R_E/d_L}$ exceeds $\approx 1$. Since $R_E \approx \sqrt{4 G M d_L}$ when $d_L \ll d_S$, the condition $\rho \lesssim 1$ corresponds to the condition $d_{L} \lesssim 4 G M(\frac{d_S}{R_S})^2$. The upper limit for the $d_L$ integral in \ref{eq:difRate_mass} is therefore $\text{min}[4 G M(\frac{d_S}{R_S})^2, d_S]$ (under the approximation that the integrand vanishes when $\rho > 1$). The integrand is $\propto v_T^4$, hence $\propto M^2 d_L^2$ by equation \ref{eq:vT}. (It should be noted that at these masses, $v_T$ is always $\ll v_c$, hence the exponential term approaches 1. We will return to this point when we consider the high-mass behavior of the curve.) Performing the integral over $d_L$ yields a term proportional to $M^2 d_L^3|_{0}^{\text{min}[4 G M(\frac{d_S}{R_S})^2, d_S]}$. In the limit that $4 G M(\frac{d_S}{R_S})^2 > d_S$, the term then scales as $M^2$, whereas in the opposite limit, the term scales as $M^5$, with the transition between these regimes occurring at $\rho \approx 1 \Rightarrow M \approx \frac{R_S^2}{4 G d_S}(\frac{d_L^*}{d_S})$, where $d_L^*$ is the average lens distance. Upon multiplying by $dn/dM \propto M^{-2}$ for our fiducial $p = 1$ UBO power law, this yields the $M^3$ low-mass behavior and $M^0$ intermediate-mass plateau seen in Figure \ref{fig:diff_rate_M_MW}. The above derivation shows that by increasing the distance to sources, the finite-source suppression can be avoided at lower lens masses. It is for this reason that the Subaru observations, with $d_S \approx 770$ kpc, can still resolve microlensing events for MW lenses without finite-source suppression at masses well below that of typical Bulge-oriented surveys with $d_S \approx 8$ kpc.

\subsection{High-mass behavior}
\label{app:highmass}

The $M^0$ dependence persists only up until the timescale of events at a particular mass  becomes shorter than the observational cadence, hence cannot be resolved. This occurs when $v_T \approx v_c$, as at larger $v_T$, events become exponentially suppressed. By the definition of $v_T$, this occurs when $2 \sqrt{4 G M d_L^*}/t_{\rm{FWHM}} \approx v_c$ where we have dropped the $\mathcal{O}(1)$  $u$-dependent term and fixed $d_L$ to its average value $d_L^*$. We approximate the integrand as zero when $v_T > v_c$, hence the lower bound on the $t_{\rm{FWHM}}$ integral becomes $\text{min}[2\sqrt{4GM d_L^*}/v_c, t_\text{min}]$. The integrand is proportional to $v_T^4 \propto M^2 / t_{\rm{FWHM}}^4$ when the exponential is $\approx 1$, hence  performing the integral over $t_{\rm{FWHM}}$ yields a term proportional to $M^2 / t^3|_{\text{min}[2\sqrt{4GM d_L^*}/v_c, t_\text{min}]}^{t_\text{max}}$. In the limit that $2\sqrt{4 GM d_L^*}/v_c < t_\text{min}$, then the expression is proportional to $M^2$, while in the opposite limit, the expression goes as $M^{1/2}$. Multiplying by the $dn/dM \propto M^{-2}$ for our fiducial $p=1$ UBO power law yields an intermediate plateau regime in which $d\Gamma/dM \propto M^0$ that transitions to high-mass $M^{-3/2}$-dependence when $v_T(t_\text{min}) \approx v_c \Rightarrow M \approx \frac{v_c^2 t_\text{min}^2}{16 G d_L^*}$. 

\begin{figure}
	\centering
	\includegraphics[width=\columnwidth]
 {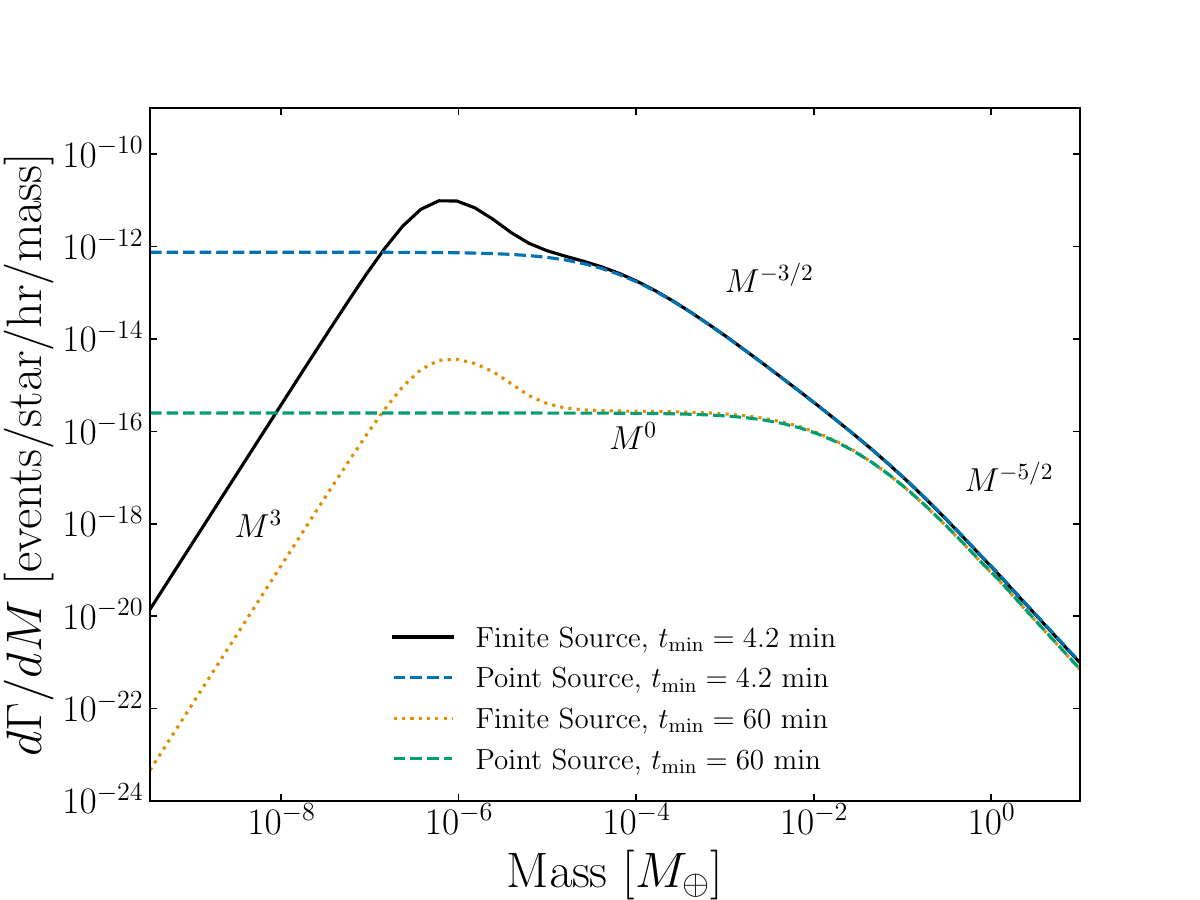}
	\caption{Differential event rate for lenses in the Milky Way disk as a function of mass. The black curve is the rate for the Subaru HSC survey, including the effects of both a finite cadence at high masses and finite-source effects at low masses. The blue dashed curve is the rate for an identical analysis but with the finite-source effects neglected. In that case, there is no low-mass loss in sensitivity. The yellow dotted curve is the event rate for a survey identical to the Subaru observation but with longer observational cadence (60 minutes). The event rate ceases to grow with decreasing mass at a higher mass, as discussed in section \ref{app:highmass}. The green dashed curve is the equivalent survey parameters to the green curve, but neglecting finite-source effects. The scaling behaviors estimated in appendix \ref{app:scalings} are shown alongside the curves.  The slight bulge in each finite-source curve near $M \approx 10^{-6}\, M_\oplus$ arises due to the narrow region of phase space in which finite-source effects lengthen observed events, allowing more short-duration events to be detected in comparison to the point-source regime.}
	\label{fig:diff_rate_M_MW}
\end{figure}

Beyond this, a similar analysis for the $d u_0$ integral yields a regime in which the  differential rate scales with $M^{-5/2}$, as can be seen in Fig. \ref{fig:diff_rate_M_MW}. This follows once again from the condition that $v_T < v_c$ to avoid exponential suppression. However, now we reinstate the $u-$dependence of $v_T$ and fix all other quantities. This yields the condition $\sqrt{u_T^2 - u_0^2} < \frac{v_c t_{\rm{FWHM}}}{4 \sqrt{G M d_L}}$. Upon a transformation to the variable $\tilde{u} \equiv \sqrt{u_T^2 - u_0^2}$, the limits of integration become $\tilde{u} \in [0, \text{min}[\frac{v_c t_{\rm{FWHM}}}{4 \sqrt{G M d_L}}, u_T]]$. In the case when the upper limit is $u_T$, the resulting integral goes as $M^0$, as we have seen before. However, when the upper limit is $\frac{v_c t_{\rm{FWHM}}}{4 \sqrt{G M d_L}}$, the leading term when $M\rightarrow \infty$ goes as $M^{-5/2}$, with the transition occurring around $M \approx v_c^2 t_\text{max}^2 / 16 G d_L u_T^2$. Note that in the geometric optics limit (applicable at these masses), $u_T \rightarrow 1$. This transition occurs at masses approximately $(t_\text{max}/t_\text{min})^2$ larger than the finite-cadence turnover, which for $t_\text{min} = 4.2$ minutes and $t_\text{max} = 3$ hours yields a ratio of $\approx 10^3$. This corresponds with the two changes in slope seen at $M\approx 10^{-4}~M_\oplus$ and $M\approx 10^{-1}~M_\oplus$ in Figure \ref{fig:diff_rate_M_MW}.

\subsection{Mass-integrated sensitivity}
\label{app:sensitivity}

In order to connect this differential event rate to a relative contribution to the expected number of events for lenses within a certain mass range, one must integrate equation \ref{eq:difRate_mass} over mass. As a result, the per dex rate of events plotted in Fig. ~\ref{fig:sensitivity} (i.e. the event rate $d\Gamma/dM$ integrated over 1 dex mass bins) displays the same overall behavior as $d\Gamma/dM$ with scalings that have an additional power of $M$. The sensitivity is therefore peaked at the cadence-limited transition point, 
\begin{equation}
\label{eq:m_turnover}
M_\text{cadence-limited} \approx \frac{v_c^2 t_\text{min}^2}{16 G d_L^*}
\end{equation}
(see section \ref{app:highmass}), which is $\mathcal{O}(10^{-4})~M_\oplus$ for a typical MW lens distance ($\approx$ 1 kpc) and dispersion ($\approx 30$ km/s) with $t_\text{min} = 4.2$ minutes \citep{niikura_microlensing_2019}. This corresponds to the peak value for the Milky Way contribution in Fig. \ref{fig:sensitivity}. It is clear from equation \ref{eq:m_turnover} that slowing the observational cadence by a single order of magnitude reduces the mass to which a survey is most sensitive by two orders of magnitude. Furthermore, the stellar dispersion within the MW disk is roughly an order of magnitude less than the Galactic bulge, which yields an additional two orders of magnitude in mass due to the $v_c^2$ dependence in equation \ref{eq:m_turnover}.

\begin{figure}
	\centering
	\includegraphics[width=\columnwidth]
 {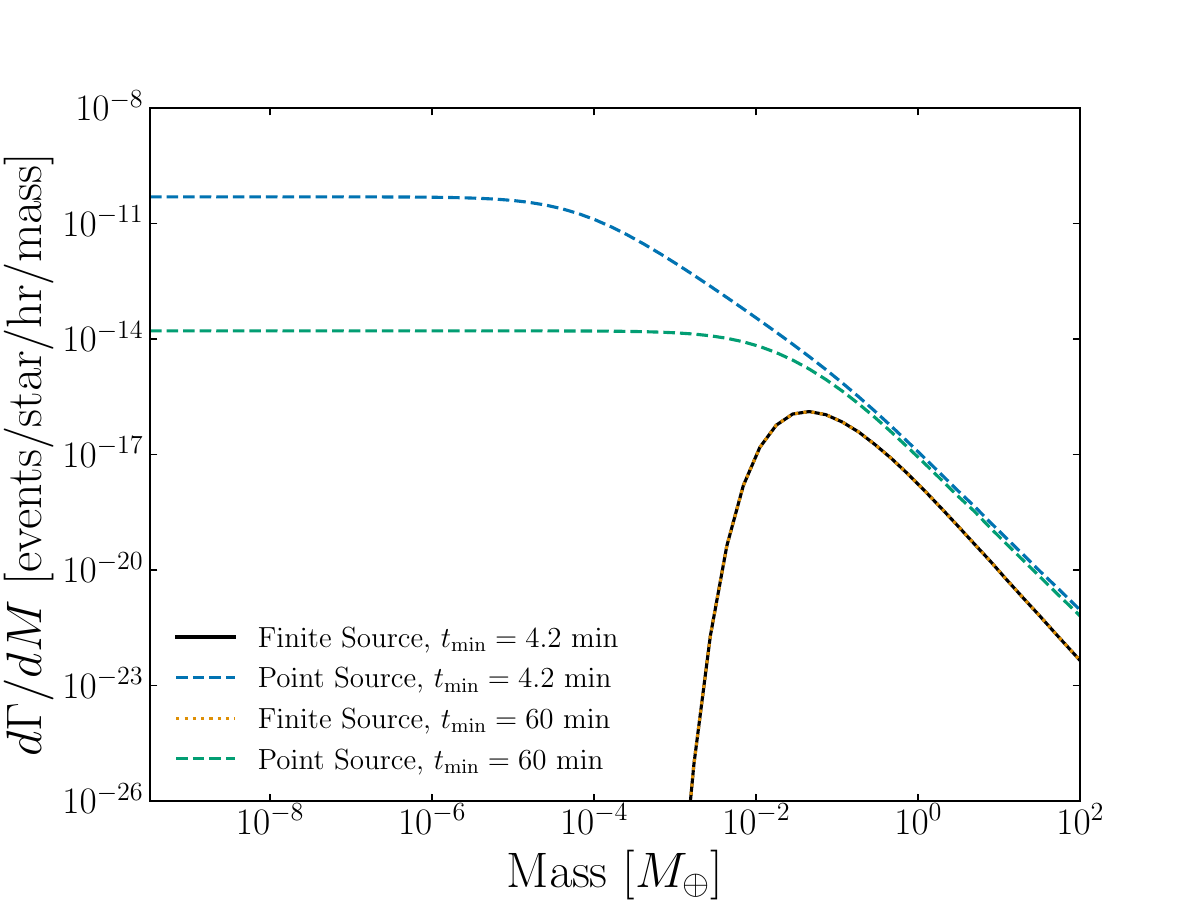}
	\caption{Same as Fig.~\ref{fig:diff_rate_M_MW} but for lenses within M31. As discussed in section \ref{app:sensitivity}, finite-source effects push the peak sensitivity to higher values of $M$ relative to MW lenses.}
	\label{fig:diff_rate_M_M31}
\end{figure}

The M31 contribution, on the other hand, is governed by the onset of finite-size effects, which become relevant at
\begin{equation}
\label{eq:m_FS}
M_\text{finite-size-limited} \approx 
\begin{cases}
\frac{R_S^2}{4 G d_S}\left(\frac{d_L^*}{d_S}\right) &d_L^* \ll d_S \\
\frac{R_S^2}{4 G d_S}\left(\frac{d_L^*}{d_S}\right)\left(1-\frac{d_L^*}{d_S}\right)^{-1} &d_L^* \approx d_S
\end{cases}
\end{equation}
(see section \ref{app:lowmass}).
It is the $d_L^* \approx d_S$ case that is relevant for M31. Upon substituting typical values in M31 for these parameters ($d_L \approx d_S \approx$ 770 kpc), one finds that the peak sensitivity occurs around $M \approx \mathcal{O}(10^{-1})~M_\oplus$, which is apparent in Figure~\ref{fig:diff_rate_M_M31}.

%%%%%%%%%%%%%%%%%%%%%%%%%%%%%%%%%%%%%%%%%%%%%%%%%%

% Don't change these lines
\bsp	% typesetting comment
\label{lastpage}
\end{document}